\documentclass[aps,pre,showpacs,preprint,superscriptaddress]{revtex4}
\usepackage{graphicx}
\usepackage{amsmath}
\usepackage{ulem}
\usepackage{bm}

\begin{document}

\title{Enhanced laser radiation pressure acceleration of protons with a gold cone-capillary}
\author{Chong Lv}
\affiliation{Key Laboratory of Beam Technology and Materials Modification of the Ministry of Education, and College
of Nuclear Science and Technology, Beijing Normal University, Beijing 100875, China}
\author{Bai-Song Xie\footnote{Corresponding author. Email address: bsxie@bnu.edu.cn}}
\affiliation{Key Laboratory of Beam Technology and Materials Modification of the Ministry of Education, and College
of Nuclear Science and Technology, Beijing Normal University, Beijing 100875, China}
\affiliation{Beijing Radiation Center, Beijing 100875, China}
\author{Feng Wan}
\author{Ya-Juan Hou}
\author{Mo-Ran Jia}
\author{Hai-Bo Sang}
\affiliation{Key Laboratory of Beam Technology and Materials Modification of the Ministry of Education, and College
of Nuclear Science and Technology, Beijing Normal University, Beijing 100875, China}
\author{Xue-Ren Hong}
\affiliation{Key Laboratory of Atomic and Molecular Physics $\&$ Functional Materials of Gansu Province, College of Physics and Electronic Engineering, Northwest Normal University, Lanzhou 730070, China}
\author{Shi-Bing Liu}
\affiliation{Strong-field and Ultrafast Photonics Lab, Institute of Laser Engineering,
Beijing University of Technology, Beijing 100124, China}
\date{\today}

\begin{abstract}
A scheme with gold cone-capillary is proposed to improve the protons acceleration and involved problems are investigated by using the two-dimensional particle-in-cell simulations. It is demonstrated that the cone-capillary can efficiently guide and collimate the protons to a longer distance and lead to a better beam quality with a dense density $\geq10n_c$, monoenergetic peak energy $E_k \sim 1.51~\mathrm{GeV}$, spatial emittance $\sim0.0088~\mathrm{mm}~\mathrm{mrad}$ with divergence angle $\theta \sim 1.0^{\circ}$ and diameter $\sim 0.5\mathrm{\mu m}$. The enhancement is mainly attributed to the focusing effect by the transverse electric field generated by the cone as well as the capillary, which can prevent greatly the protons from expanding in the transverse direction. Comparable to without the capillary, the protons energy spectra have a stable monoenergetic peak and divergence angle near to $1.0^{\circ}$ in longer time. Besides, the efficiency of acceleration depending on the capillary length is explored, and the optimal capillary length is also achieved. Such a target may be benefit to many applications such as ions fast ignition in inertial fusion, proton therapy in medicine and so on.
\end{abstract}
\maketitle

\section{Introduction}

With the wide application, the charged particles acceleration based on the laser-plasma interaction have been rapidly concerned by many simulations and experiments \cite{Schwoerer,Esarey,Corde}. In recent years, the proton acceleration generated by an ultraintense laser irradiating a solid target has been investigated extensively due to the potential applications in ions fast ignition (FI) of inertial confinement fusion (ICF) \cite{Roth,Temporal,Naumova}, compact proton sources for cancer therapy \cite{Bulanov,Ledingham}, laboratory astrophysics \cite{Remington} and so on. For instance, the FI by proton beams to transport a few of meters to the dense core usually requires a high beam quality \cite{Roth}. However, the coupling efficiency from laser to protons is too low to achieve the enough ignition energy. Therefore, how to gain a high quality beam with high energy, low energy spread and small size is a challenge topic currently.

 In the past years, several acceleration schemes have been proposed and studied for obtaining higher quality proton beams by many researchers. According to the parameters variety of  the laser and targets, the acceleration approaches are usually classified into target normal sheath acceleration (TNSA) \cite{Snavely,Fuchs,Ma,Bake}, breakout afterburner acceleration (BOA) \cite{Yin,Yin1,Flippo}, shock wave acceleration \cite{Silva,Wei}, and radiation pressure acceleration (RPA) \cite{Esirkepov,Yan,Qiao,Hong,Hong2016}, etc.. The TNSA, which has been studied rapidly and implemented in the experiments, is that protons can be accelerated to high energy through the electrostatic sheath field created by the hot electrons expanding into vacuum at the rear side of the target. Nevertheless, the practical application is limited because of the obtained protons possessing a large divergence, large energy spread and low number density.

Moreover, in order to obtain a dense and monoenergetic beam, the RPA is proposed to efficiently accelerate the protons. This scheme is mainly associated to the intense space-charge field, which is created by the radiation pressure of the laser. Yet, multi dimensional simulations show that the RPA is not enough stable due to the excited transverse instability, such as Weibel instability and Rayleigh-Taylor-like instability \cite{Pegoraro,Yan1,Palmer}. In order to suppress the instability, many solutions are presented and studied in detail like of with a properly tailored laser \cite{Pegoraro,Yin2} or/and with a shaped foil target \cite{Chen,Yu}, etc.. Recently Zou \textit{et al.} \cite{Zou} proposed first a cone target used to focus laser and guide fast electrons in FI \cite{Kodama}. While the cone target can suppress the transverse expansion by a co-moving focusing electric field and achieve a better beam quality, however, the proton beams will diffuse following the laser field when they fly out from the right exit of the cone. This deficiency makes the protons not be accelerated availably to a farther distance.

In this work, a gold cone with a capillary attached to the behind is proposed and involved problems are studied by using the two-dimensional particle-in-cell (PIC) simulations. The results show that this cone-capillary target can availably collimate and guide the protons, then achieve a better beam quality with a density $\geq10n_c$, monoenergetic peak $E_k \sim 1.51~\mathrm{GeV}$, spatial emittance $\sim0.0088~\mathrm{mm~mrad}$ with divergence angle $\theta \sim 1.0^{\circ}$ and diameter $\sim 0.5\mathrm{\mu m}$  in a farther distance. The enhanced reasons are mainly attributed to the focus by the transverse electric field $E_y$ generated in the cone as well as the capillary. Besides, in the process of exploring the capillary parameters, one finds that the capillary lengths act as a important role in the protons acceleration.

This paper is organized as follows. Section II outlines the target configurations, simulation parameters and results. Besides, the enhanced reasons of protons acceleration are also discussed and analyzed in detail. The capillary parameters are examined in Section III , and the optimal capillary length is given. Finally a brief summary is presented in Sec. IV.

\section{NUMERICAL SIMULATION AND RESULTS}

For the purpose to demonstrate the enhancement of proton acceleration, we have designed two kinds of targets by using the gold cone and gold cone-capillary. We have investigated the created protons quality by them via the two-dimensional PIC simulations.

\subsection{Target configurations and simulation parameters}

\begin{figure}[htbp]\suppressfloats
\includegraphics[width=15cm]{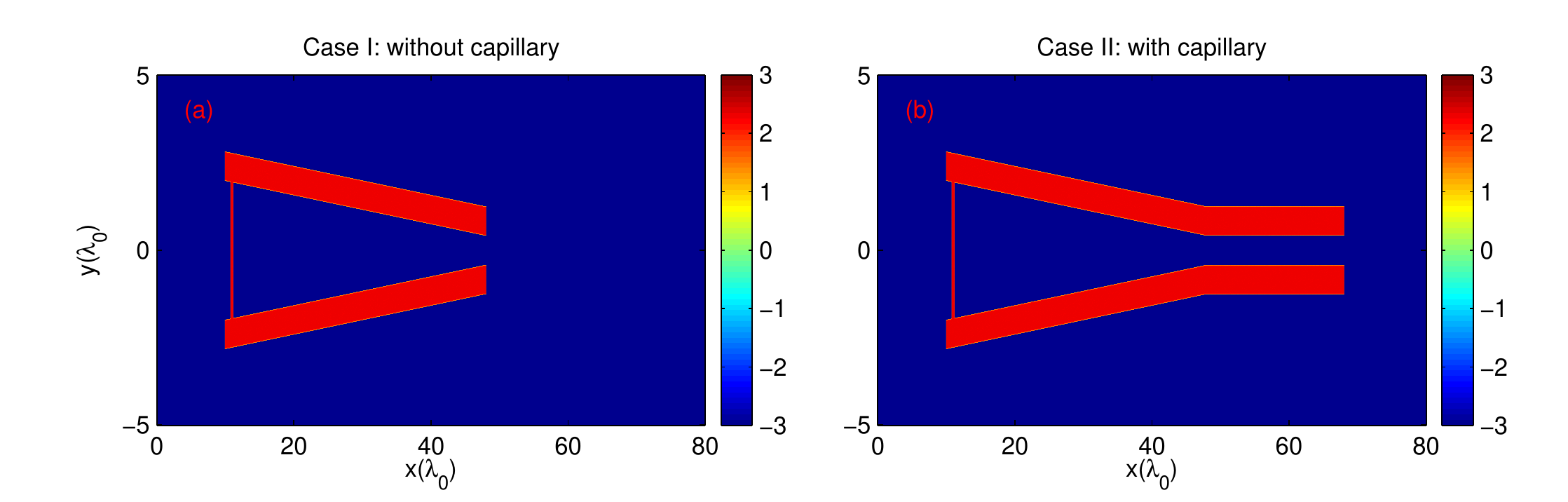}
\caption{(Color online) Simulation box and initial target structure. Initial plasma density on a logarithmic scale of the gold cone for case I in (a) and the gold cone-capillary for case II in (b), respectively. Here the density is normalized by the critical density $n_c$.}
\label{fig:1}
\end{figure}

Two different cases of targets are illustrated in Fig. \ref{fig:1} whose the case II by Fig. \ref{fig:1}(b) represents the gold cone with a capillary attached behind the cone and as a comparison the case I by Fig. \ref{fig:1}(a) stands for the gold cone only.

The simulation is performed with the PIC code EPOCH \cite{Arber}. The simulation box is $80\lambda_0 \times 10\lambda_0$ and the grid size is $\Delta x = \Delta y = 0.02 \lambda_0$ with $4000 \times 500$ grid cells. The preionized cone or/and cone-capillary consist of electrons and Au ions whose density is established to be $n_{Au}=100n_c$. The cone is located from $x=10\lambda_0$ to $x=48\lambda_0$ with the cone-angle of $4.73^{\circ}$ from horizon. The thickness of its wall is $0.8\lambda_0$ and the diameter of the left opening is $L=4\lambda_0$. A proton-rich foil, which is full ionized with density $n_H=100n_c$ and thick $l=0.4\lambda_0$, is situated at a distance $W=0.8\lambda_0$ away from the left boundary of the cone. A capillary with length $d_0=20\lambda_0$ is attached behind the cone. Here $\lambda=1.0 ~\mathrm{\mu m}$ is the laser wavelength and $n_c = m_e\omega^2/4\pi e^2$ ($\omega$ laser frequency, $m_e$ electron mass and $-e$ electron charge) is the critical electron plasma density. Besides, each cell consists of $49$ weighted particles per species in our simulations.

A circular polarized (CP) laser pulse is normally incident from the left boundary of the box. The laser has a peak intensity of $I = 2.74 \times 10^{22} ~ \mathrm{W/cm^2}$, and rises to the maximum value in $T_0$ then remains constant for $9T_0$, where $T_0$ is the laser period. The intensity profile of the laser is Gaussian with spot size of $4.0 ~ \mathrm{\mu m}$ [full width at half maximum (FWHM)]. Besides, we use the absorption boundary in $x$ direction and $y$ direction for both fields and particles, respectively.

\subsection{Simulation results}

\begin{figure}[htbp]\suppressfloats
\includegraphics[width=15cm]{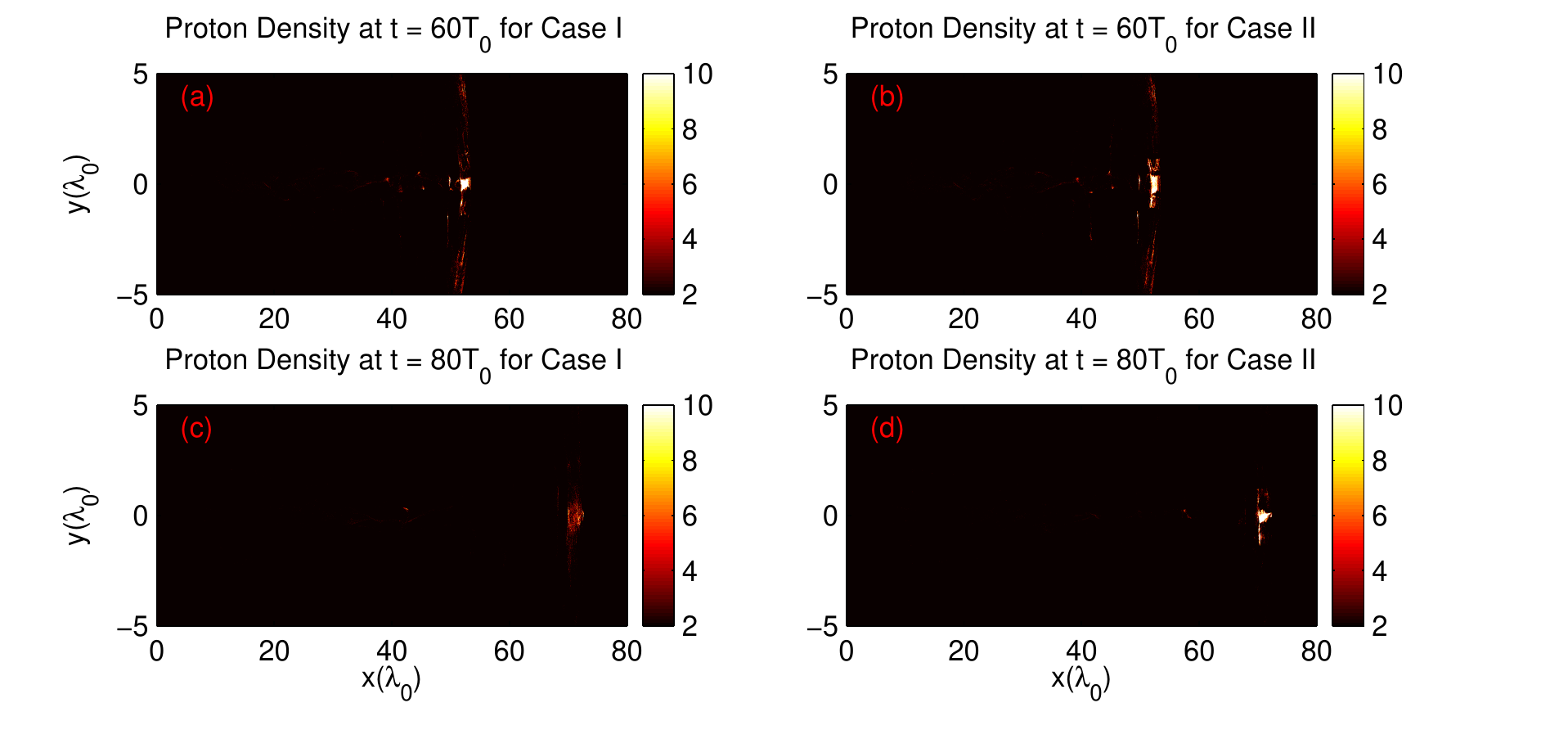}
\caption{(Color online) The proton density of the case I [(a) and (c)] and the case II [(b) and (d)] at $t=60T_0$ [(a) and (b)] and $80T_0$ [(c) and (d)]. Here the proton density is normalized by $n_c$.}
\label{fig:2}
\end{figure}

\begin{figure}[htbp]\suppressfloats
\includegraphics[width=15cm]{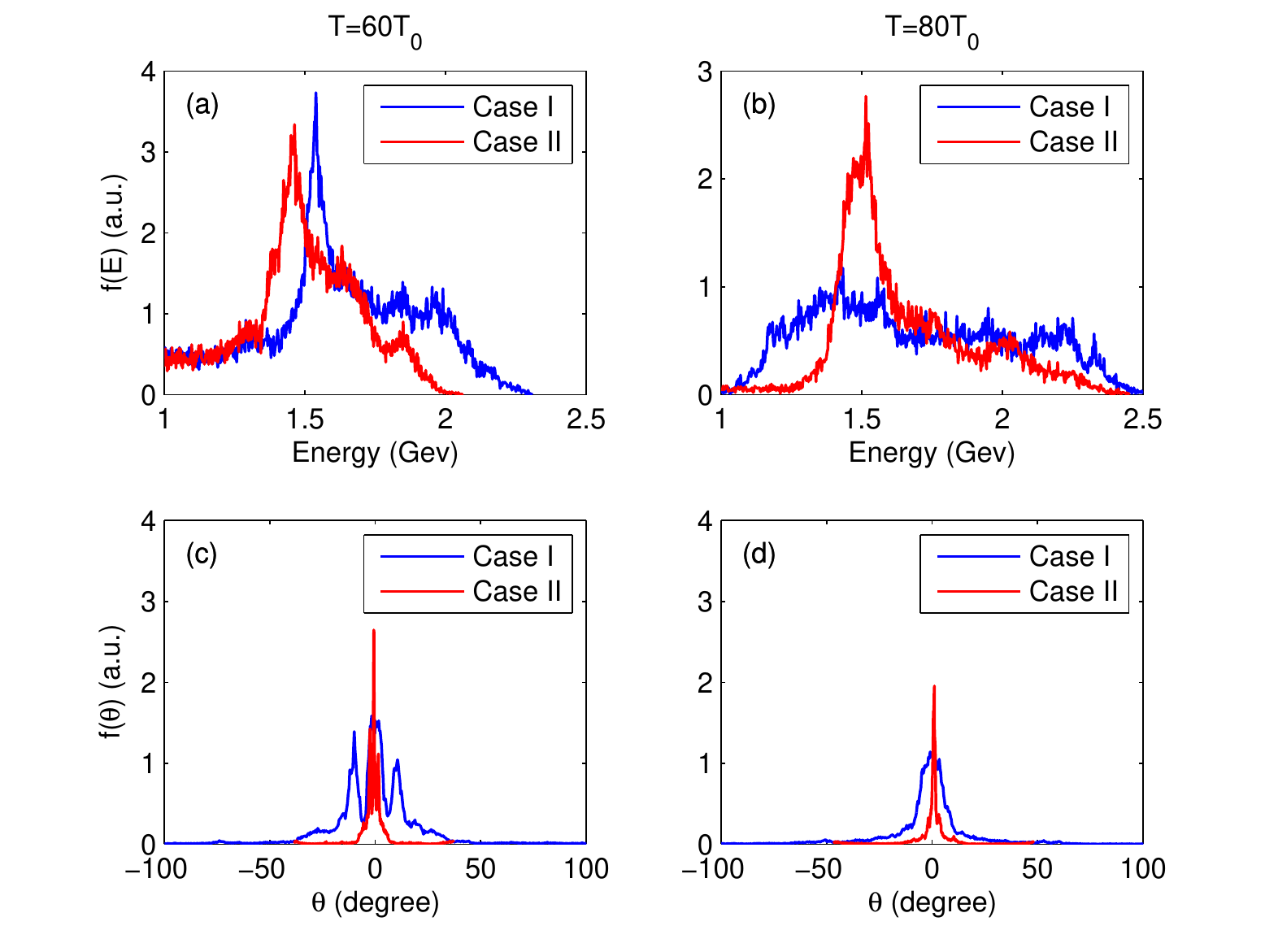}
\caption{(Color online) The proton energy spectra and angular distribution for both cases at $t=60T_0$ [(a), (c)] and $t=80T_0$ [(b), (d)]. Blue and red lines represent the case I and the case II, respectively.}
\label{fig:3}
\end{figure}

Now let us to see the results of proton acceleration in the two different kinds of targets. Figure \ref{fig:2} shows the distributions of the proton-rich foil density in both cases. At $t=60T_0$, the transverse expansion of the proton-rich foil is suppressed greatly in both cases, which is good for the transverse collimation in the acceleration process, as shown in Fig. \ref{fig:2}(a) and \ref{fig:2}(b). These results are in accordance with the Ref. \cite{Zou}. Meanwhile in comparison with the case I and II the proton density is more compacter for the case I. But as time goes on the situation is reversed since proton density becomes more and more compact for the case II. This point can be seen at $t=80T_0$ in Fig. \ref{fig:2}(c) and (d) that a dense proton source can be gained with a density $\geq10n_c$ and bunch transverse diameter $\leq0.5\lambda_0$ in case II, but in contrast the proton density decreases significantly and bunch also diffuses seriously in the transverse direction in the case I.

To demonstrate the enhanced acceleration effect by using the cone-capillary, the energy spectra and angular distribution of the protons are shown in Fig. \ref{fig:3}.

For the energy spectra they are plotted in Fig. \ref{fig:3}(a) and \ref{fig:3}(b) at different time. It shows that there are a monoenergetic proton bunches in both cases at $t=60T_0$. Similarly as in the density the better quality seems to be got in case I at this relative earlier time. For example, see Fig. \ref{fig:3}(a), the bunch quality of a peak energy $E_k \sim 1.53~\mathrm{GeV}$ with energy spread $\sim 5\%$ in the case I is indeed better than that of a peak energy $E_k \sim 1.47~\mathrm{GeV}$ with energy spread $\sim 12\%$ in the case II. However with time increasing, the situation is reversed again for the energy spectra. In fact the monoenergetic peak structure disappears evenly in the case I at $t=80T_0$ while it remains a good quality by a peak energy $E_k \sim 1.51~\mathrm{GeV}$ with energy spread $\sim 7\%$ in the case II, see Fig. \ref{fig:3}(a), even if the protons have moved out from the capillary. On the other hand the protons number at the monoenergetic peak in the case II is almost three times higher than that in the case I.

For the angular distribution they are plotted in Fig. \ref{fig:3}(c) and \ref{fig:3}(d). The momentum angle of protons is defined by $\theta=\tan^{-1}(p_y/p_x)$, where $p_y$ and $p_x$ are the momentum of the protons in the transverse $y$-axis and longitudinal $x$-axis, respectively. In the case II, it is surprising to see that the FWHM of the proton divergence distribution maintains $1.0^{\circ}$ so that the spatial emittance is about $0.0088~\mathrm{mm}~\mathrm{mrad}$  for the bunch of $0.5\mathrm{\mu m}$ diameter at time of $T=80T_0$, which is greatly beneficial for the practical applications. In the case I, however, the proton divergence angle is large, which decreases slightly from about $20.0^{\circ}$ at $60T_0$ to $14.0^{\circ}$ at $80T_0$. Evenly the corresponding spatial emittance increases from $0.176$ at $60T_0$ to $0.246~\mathrm{mm}~\mathrm{ mrad}$ at $80T_0$.

These results obviously indicate that the case II is benefit to accelerate the protons to a farther distance and get a improved high quality protons bunch. Thus it is necessary to clarify what is behind this improvement. We will make some analysis and discussion on the electric and magnetic fields for the both cases in the next subsection, which will show their important roles, in particular the role by the transverse electric field, on the enhancement of proton quality.

\subsection{Analysis and discussion}

\begin{figure}[htbp]\suppressfloats
\includegraphics[width=15cm]{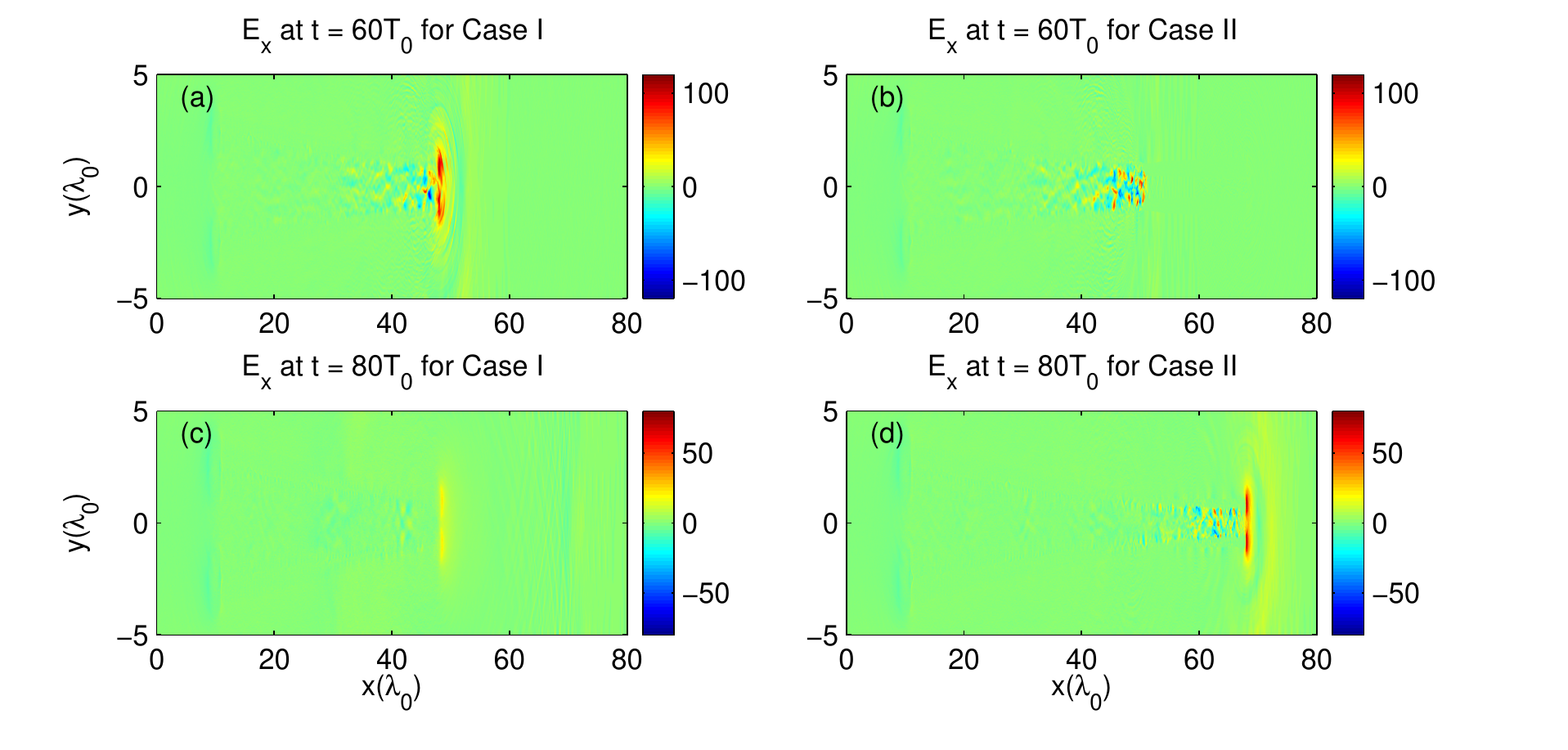}
\caption{(Color online) The contours of  longitudinal electric fields $E_x$ for the case I [(a) and (c)] and case II [(b) and (d)] at $60T_0$ [(a) and (b)] and $80T_0$ [(c) and (d)]. Here the electric field is normalized by $m_e\omega c/e$.}
\label{fig:4}
\end{figure}

\begin{figure}[htbp]\suppressfloats
\includegraphics[width=15cm]{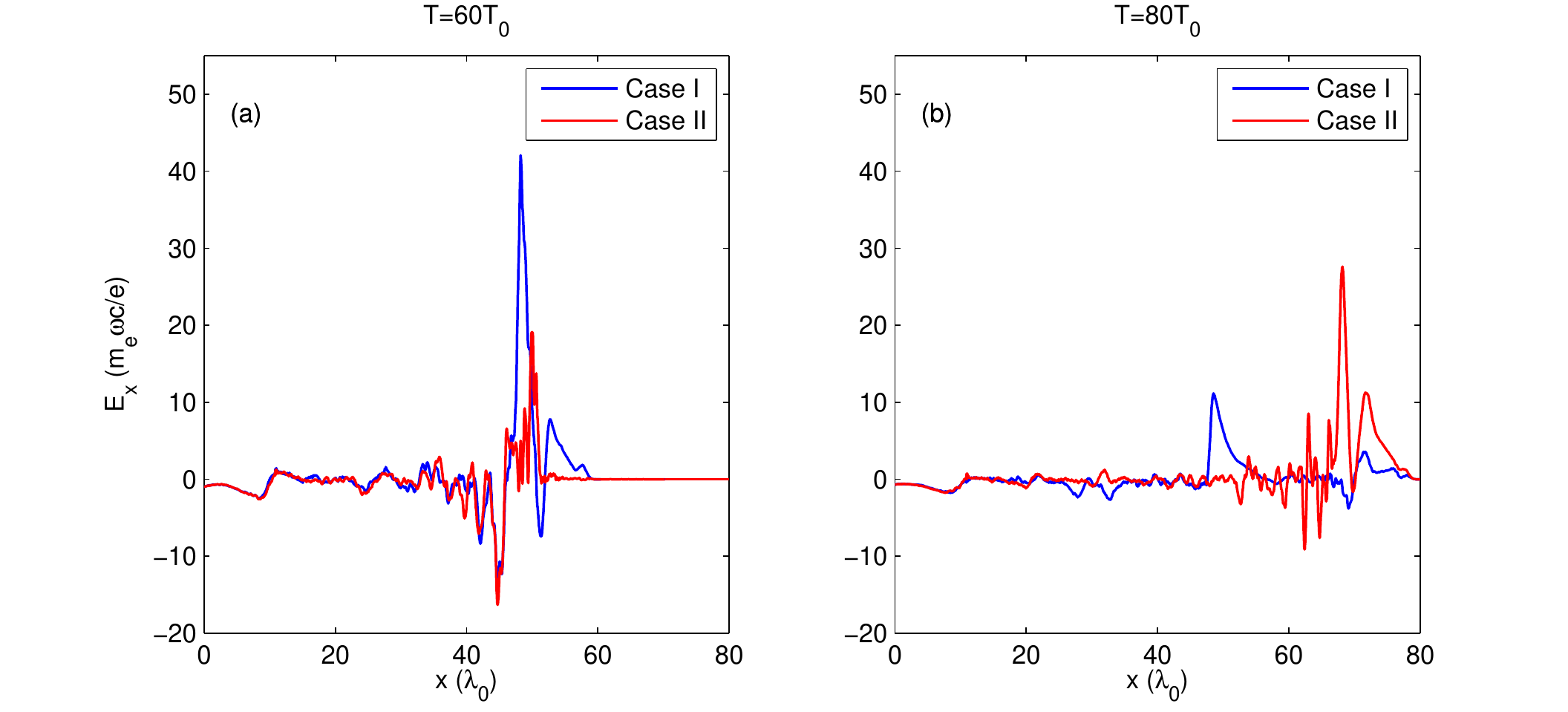}
\caption{(Color online) The slices of  longitudinal electric fields $E_x$ at $60T_0$ (a) and $80T_0$ (b) in the position $y=0$, respectively. Blue and red lines represent the case I and the case II, respectively. Here the electric field is normalized by $m_e\omega c/e$.}
\label{fig:5}
\end{figure}

In the above subsection we have exhibited the enhanced acceleration effect of protons by using the cone-capillary. Now let us to see how this enhancement is affected mainly by the electric and magnetic fields in different target structures which plays a key role to the improved collimation of produced protons bunch.

The longitudinal electric fields $E_x$  at $t=60T_0$ and $80T_0$ are plotted respectively in Fig. \ref{fig:4}. At $t=60T_0$ [Fig. \ref{fig:4}(a) and \ref{fig:4}(b)],  in the region of about $51\lambda_0<x<57\lambda_0$, one can note that the $E_x$ for the case I is obviously stronger than that for the case II. Thus, the protons can be accelerated to higher energy, corresponding to a higher monoenergetic peak structure in the case I. But as time goes on [Fig. \ref{fig:4}(c) and \ref{fig:4}(d)], the $E_x$  diffuses along transverse and becomes quite weak in the region of about $71\lambda_0<x<77\lambda_0$. However what is interesting for the case II, a stronger $E_x$ is built in a farther distance due to the focus of  the capillary at $t=80T_0$, which can provide driving force for the proton acceleration in long time. In order to show this feature clearer, the slices of the  $E_x$ in position of $y=0$ are plotted in Fig. \ref{fig:5} for both cases. One can see that, for the case I, there exists a stronger $E_x$ to  accelerate the protons forward at early time, while  decreases significantly at later time, see the blue lines in Fig. \ref{fig:5}(a) and \ref{fig:5}(b). For the case II, however, the $E_x$ becomes stronger and stronger with time increasing, see the red lines in Fig. \ref{fig:5}(a) and \ref{fig:5}(b), which is prevailing to the protons acceleration in longer time because of the focus by the capillary.

\begin{figure}[htbp]\suppressfloats
\includegraphics[width=15cm]{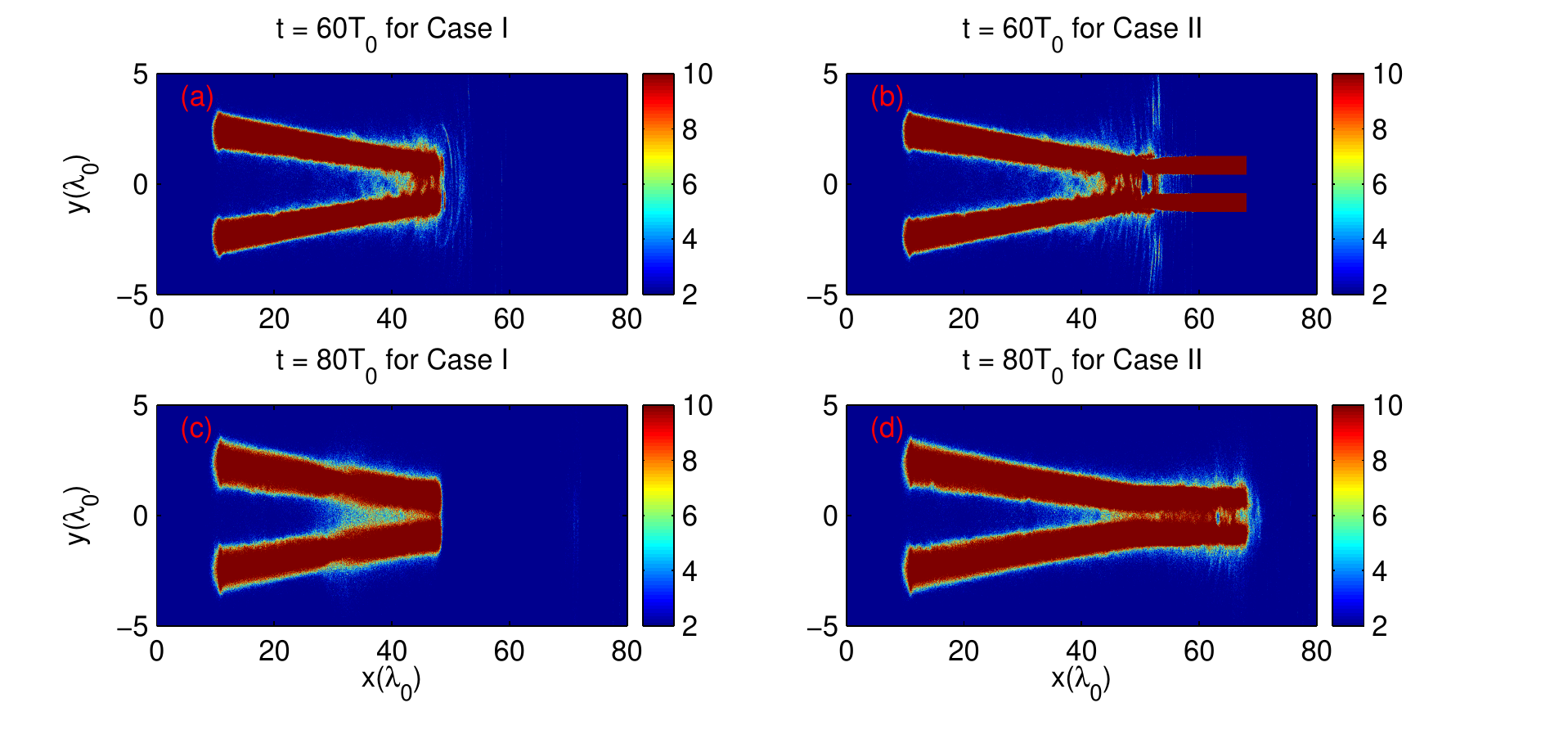}
\caption{(Color online) The electron density of the case I [(a) and (c)] and the case II [(b) and (d)] at $t=60T_0$ [(a) and (b)] and $80T_0$ [(c) and (d)]. Here the electron density is normalized by $n_c$.}
\label{fig:6}
\end{figure}

\begin{figure}[htbp]\suppressfloats
\includegraphics[width=15cm]{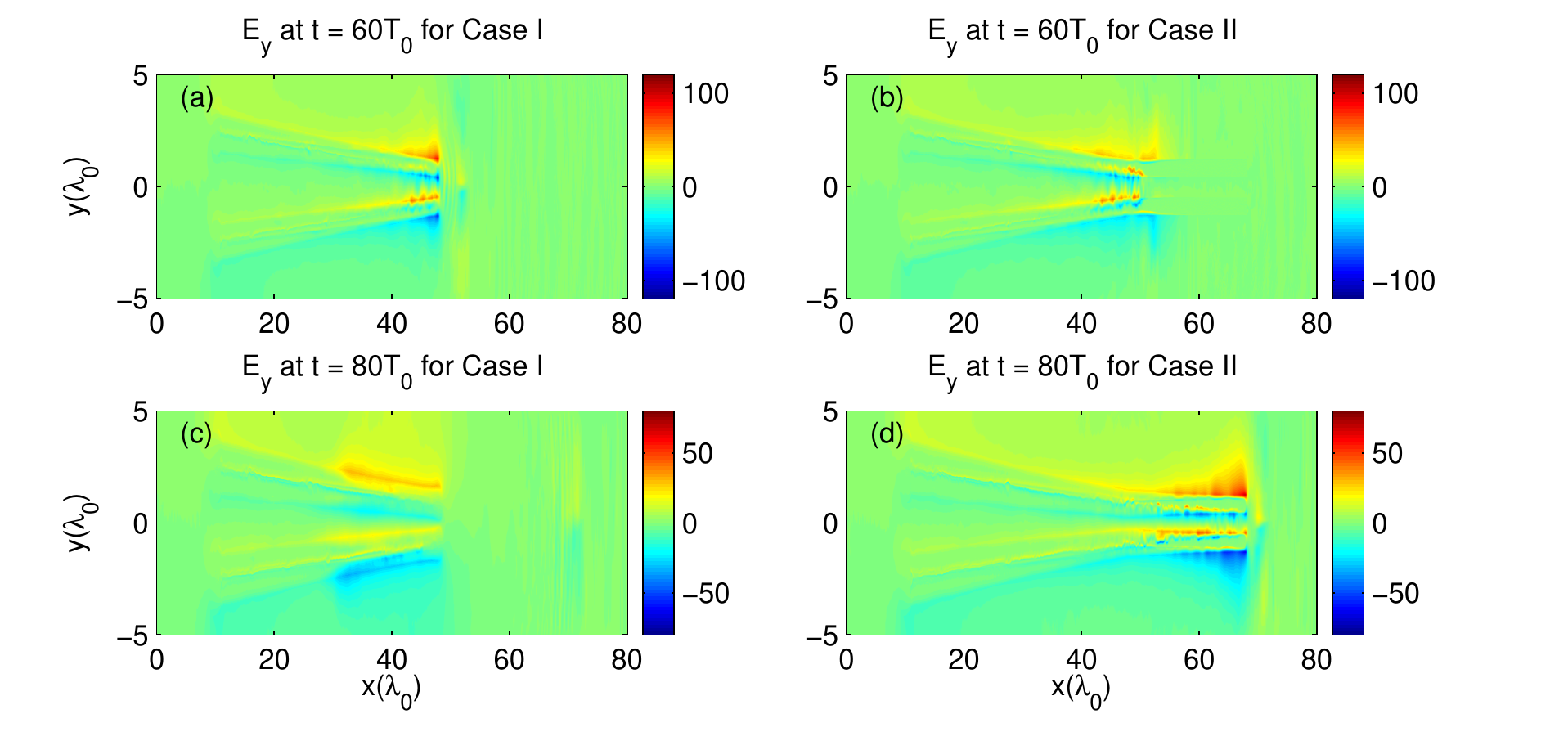}
\caption{(Color online) The contours of  transverse electric fields $E_y$ for the case I [(a) and (c)] and case II [(b) and (d)] at $60T_0$ [(a) and (b)] and $80T_0$ [(c) and (d)]. Here the electric field is normalized by $m_e\omega c/e$.}
\label{fig:7}
\end{figure}

\begin{figure}[htbp]\suppressfloats
\includegraphics[width=15cm]{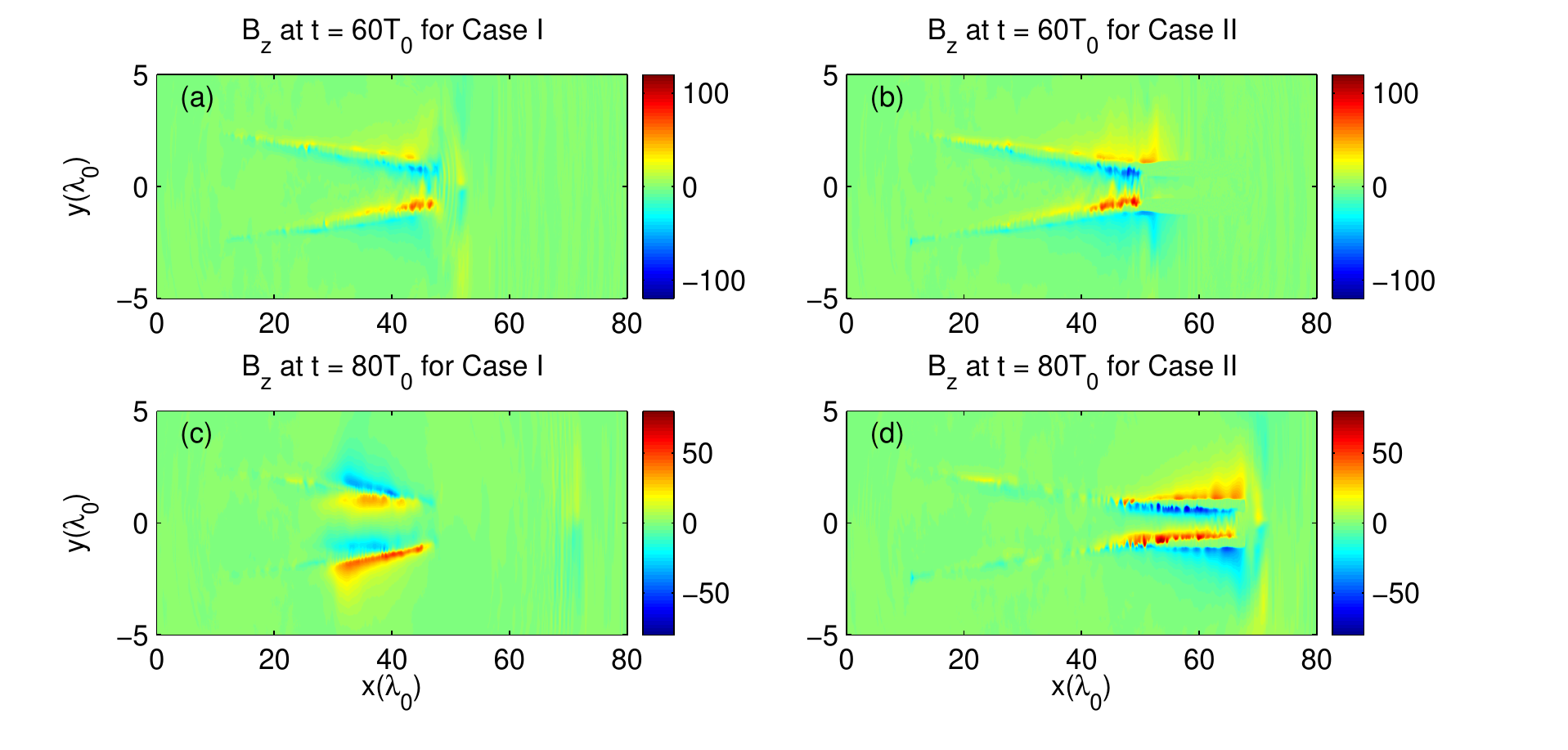}
\caption{(Color online) The contours of  quasistatic magnetic fields $B_z$ for the case I [(a) and (c)] and case II [(b) and (d)] at $60T_0$ [(a) and (b)] and $80T_0$ [(c) and (d)]. Here the magnetic field is normalized by $m_e\omega c/e$.}
\label{fig:8}
\end{figure}

When the laser propagates into the cone or/and cone-capillary, surface electrons are first pulled out into the vacuum by laser electric fields, then they are accelerated forward by the Lorentz force of $\bm{v} \times \bm{B}$, as shown in Fig. \ref{fig:6}. The electrons expanded into the vacuum continue to propagate along the cone wall and induce a strong transverse electric field $E_y$, shown in Fig. \ref{fig:7}, and meanwhile an intense quasistatic magnetic field $B_z$, shown in Fig. \ref{fig:8}. The induced quasistatic magnetic field plays a role of driving the electrons out from the cone surface, while the $E_y$ acts to draw electrons back into the inside of the cone. Moreover, it is worth to note that the moving $E_y$ can also play the part of focusing the protons in the transverse, which has been proposed by Zou \textit{et al.} \cite{Zou}. This results in a balance between the electric field and the magnetic field \cite{Kodama1,Ma1} that leads to the collimation and guidance of the electrons along the cone or/and cone-capillary. At $t=60T_0$, the protons can collimated by the transverse electric field $E_y$ for both cases, as shown in Fig. \ref{fig:7}(a) and \ref{fig:7}(b). However, as time goes on, the $E_y$ disappears in the region of $48\lambda_0<x<68\lambda_0$ for the case I. While it is still quite strong due to the existence of the capillary for the case II, which ensures to focus sustainedly the protons in a farther distance.

\section{EFFECT OF THE TARGET PARAMETERS ON THE PROTONS ACCELERATION}

From the preceding discussion, we have researched how the protons acceleration is enhanced by the cone-capillary. One can conclude that the $E_y$ generated by the capillary acts as the crucial role in guiding and collimating the protons. Accordingly, the length of capillary is the key parameter in dominating the beam quality. This reminds us to make more simulations to see whether there exists an optimal capillary length for protons quality when the other parameters are fixed.

\begin{figure}[htbp]\suppressfloats
\includegraphics[width=15cm]{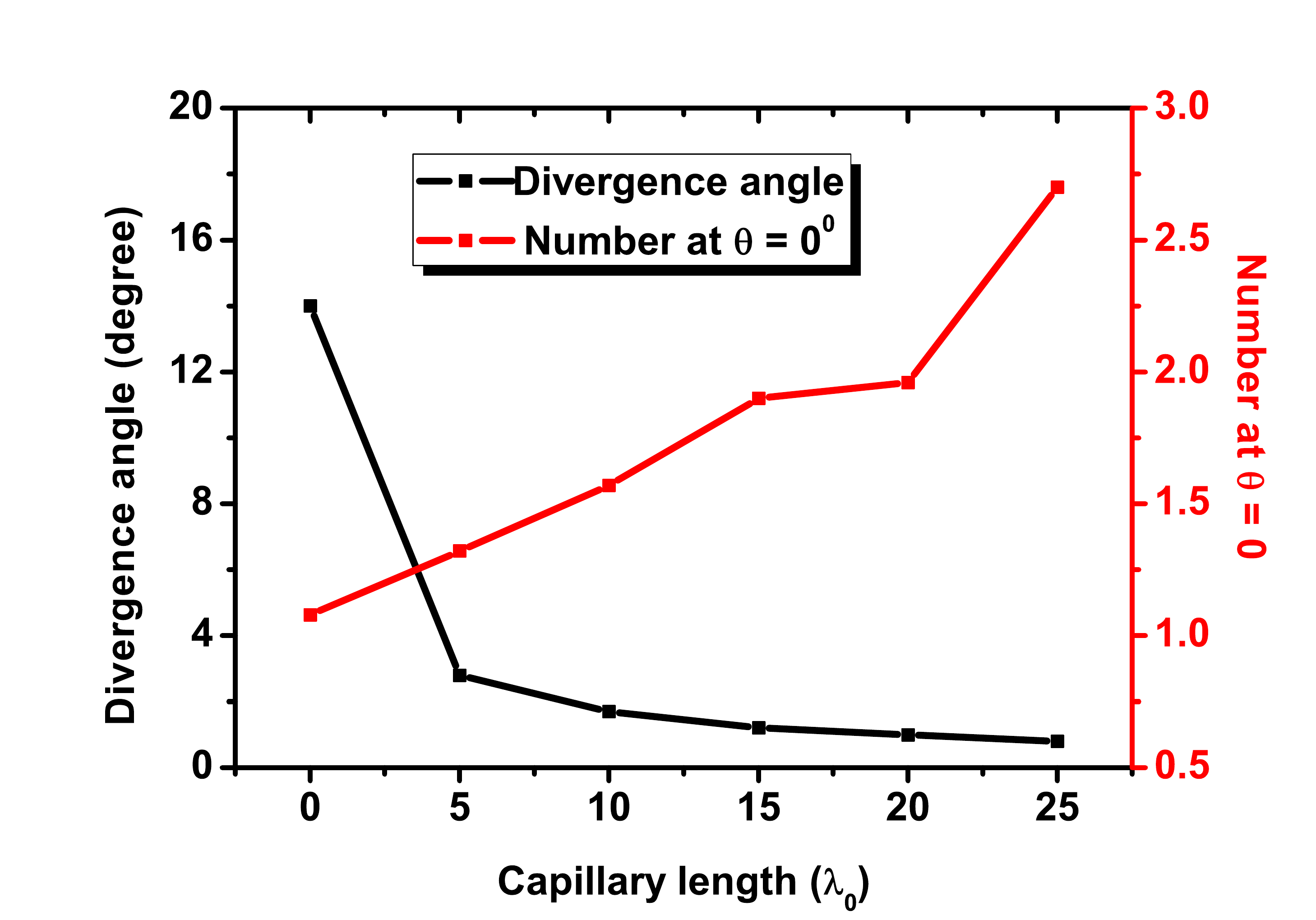}
\caption{(Color online) The divergence angle $\theta$ (black curve) and the number density locating the peak (red curve) as the function of the length $d_0$ of the capillary, respectively.}
\label{fig:9}
\end{figure}

\begin{figure}[htbp]\suppressfloats
\includegraphics[width=15cm]{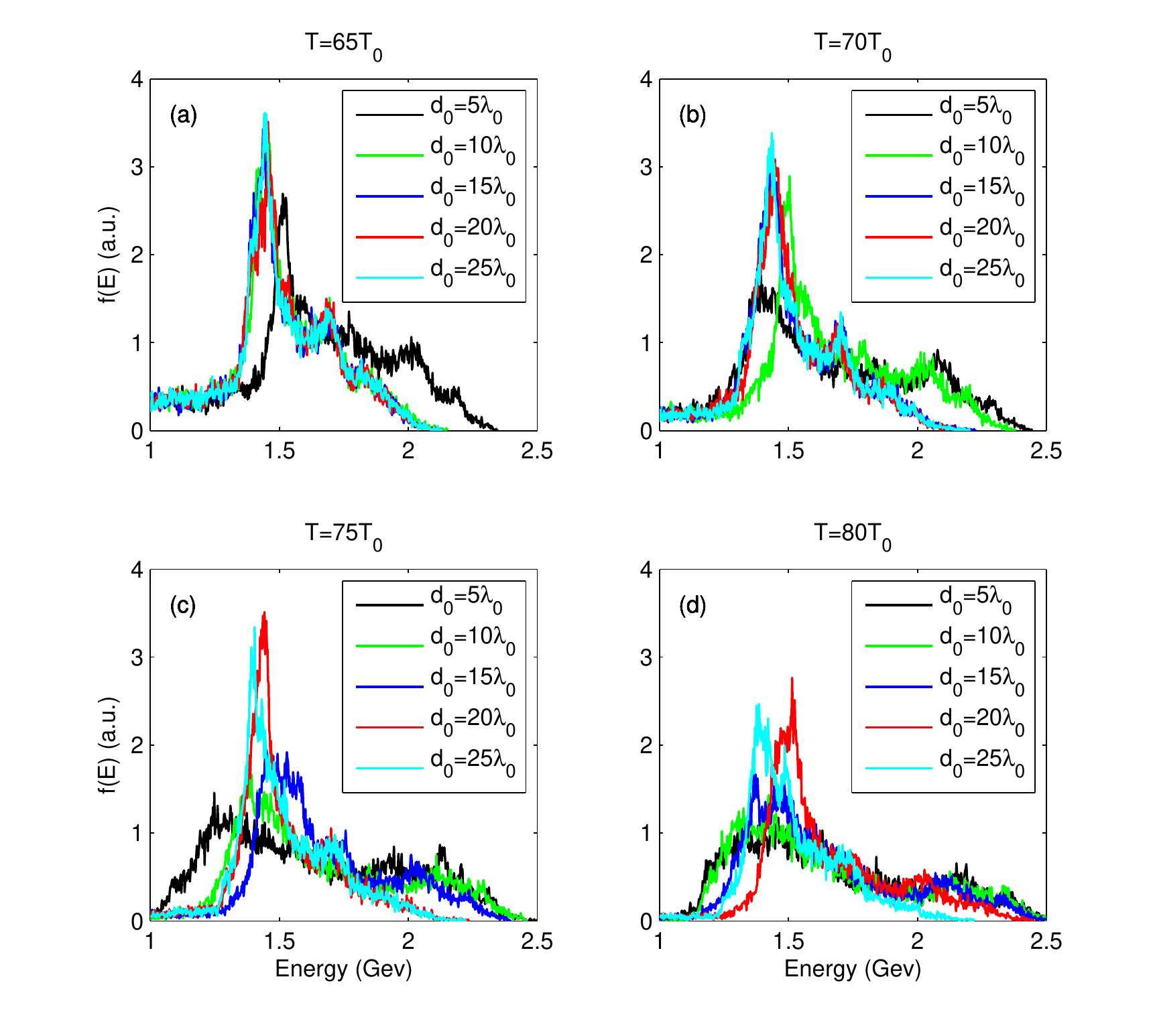}
\caption{(Color online) The proton energy spectra for different length of the capillary at (a)  $t=65T_0$, (b)  $t=70T_0$, (c)  $t=75T_0$ and (d)  $t=80T_0$, respectively.}
\label{fig:10}
\end{figure}

The proton divergence angle $\theta$ and number density of the peak energy are plotted as the function of the capillary length $d_0$, as shown in Fig. \ref{fig:9}. Obviously the length $d_0=0$ represents the case I. It shows that the $\theta$ decreases with $d_0$ , see the black curve in Fig. \ref{fig:9}, and the number density around of the peak energy increases with $d_0$, see the red curve in Fig. \ref{fig:9}. Certainly these results are mainly attributed to the collimation of the $E_y$ generated by the capillary. On the other hand, it should be emphasized that the number density increases appreciably when $d_0=25\lambda_0$. We believe the reason is that the bulk protons have not escaped out from the right opening of the capillary.

Moreover, Fig. \ref{fig:10} exhibits the proton energy spectra with different capillary lengths at different moments. As time goes on, it shows that the shorter the length of the capillary, the earlier the monoenergetic peak of the proton beam disappears. Thus this can be summed up that the shorter capillary, for example $d_0=5\lambda_0$, $10\lambda_0$ and $15\lambda_0$, is not conducive to accelerating protons to a farther distance. We can also note that for the longer capillary such as $d_0=20\lambda_0$ and $25\lambda_0$,  the monoenergetic peak are approximately coincident at early time, however, at later time , see Fig. \ref{fig:10}(d), comparing the cases of $d_0=20\lambda_0$ with the length $d_0=25\lambda_0$, the monoenergetic peak in both of energy value and number density are better for the former case. This is not surprising because that the longer capillary would make the longitudinal electric field $E_x$ decreasing, which would result in the decrease of the total conversion efficiency from the laser to protons. Overall, the extensive simulation results indicate an optimal proper capillary length existence, e.g. $d_0=20\lambda_0$ shown in Fig. \ref{fig:10}, by which one can achieve a high quality proton source even in longer time.

\section{Summary}

In summary, we have investigated the enhanced proton acceleration in the cone-capillary by using the 2D3V PIC simulations performance. Compared with the cone without the capillary, our results show that the protons can be accelerated and guided to a farther distance by using the cone-capillary. This enhancement has been analyzed and discussed in detail. First, when the protons enter into the cone,  they can be collimated in transverse direction by the $E_y$ generated in the cone for both cases. Afterwards, they can be continually accelerated and guided by the $E_y$ generated in the capillary, which results in a higher quality proton source in longer time for the case II. Yet it decreases significantly and diffuses along transverse direction for the case I. As a result, a dense proton source can be achieved with a density $\geq10n_c$, monoenergetic peak $E_k \sim 1.51~\mathrm{GeV}$, spatial emittance $\sim0.0088~\mathrm{mm}~\mathrm{mrad}$ with divergence angle $\theta \sim 1.0^{\circ}$ and diameter $\leq0.5\mathrm{\mu m}$ by the cone-capillary. Lastly, the capillary lengths are also optimized and analyzed. These results may have many important implications such as the ions fast ignition, the medical applications as well as the laboratory astrophysics and so on.

\section{Acknowledgements}

This work was supported by the National Natural Science Foundation of China (NSFC) under Grant Nos. 11475026, 11305010, 11365020 and the NSAF of China under Grant No. U1530153. The computation was carried out at the HSCC of the Beijing Normal University. The authors are particularly grateful to CFSA at University of Warwick for allowing us to use the EPOCH.


\begin{thebibliography}{99}\suppressfloats
	
	\bibitem{Schwoerer}
	H. Schwoerer, S. Pfotenhauer, O. J\"{a}ckel, K-U Amthor, B. Liesfeld, W. Ziegler, R. Sauerbrey, KWD Ledingham, and T. Esirkepov, Nature (London) \textbf{439}, 455 (2006).
	
	\bibitem{Esarey}
	E. Esarey, C. B. Schroeder, and W. P. Leemans, Rev. Mod. Phys.
	\textbf{81}, 1229 (2009).
	
	\bibitem{Corde}
	S. Corde, K. Ta Phuoc, G. Lambert, R. Fitour, V. Malka, A. Rousse, A. Beck, and E. Lefebvre, Rev. Mod. Phys. \textbf{85}, 1 (2013).
	
	\bibitem{Roth}
	M. Roth, T. E. Cowan, M. H. Key, S. P. Hatchett, C. Brown, W. Fountain,
	J. Johnson, D. M. Pennington, R. A. Snavely, S. C. Wilks, K. Yasuike, H.
	Ruhl, F. Pegoraro, S. V. Bulanov, E. M. Campbell, M. D. Perry, and H.
	Powell, Phys. Rev. Lett. \textbf{86}, 436 (2001).
	
	\bibitem{Temporal}
	M. Temporal, J. J. Honrubia, and S. Atzeni, Phys. Plasmas \textbf{9}, 3098 (2002).
	
	\bibitem{Naumova}
	N. Naumova, T. Schlegel, V. T. Tikhonchuk, C. Labaune, I. V. Sokolov,
	and G. Mourou, Phys. Rev. Lett. \textbf{102}, 025002 (2009).
	
	\bibitem{Bulanov}
	S. V. Bulanov, and V. S. Khoroshkov, Plasma Phys. Rep. \textbf{28}, 453 (2002).
	
	\bibitem{Ledingham}
	K. W. D. Ledingham, W. Galster, and R. Sauerbrey, Br. J. Radiol. \textbf{80}, 855 (2007).
	
	\bibitem{Remington}
	B. A. Remington, D. Arnett, R. P. Drake, and H. Takabe, Science \textbf{284}, 1488 (1999).
	
	\bibitem{Snavely}
	R. A. Snavely, M. H. Key, S. P. Hatchett, T. E. Cowan, M. Roth,
	T. W. Phillips, M. A. Stoyer, E. A. Henry, T. C. Sangster, M. S. Singh,
	S. C. Wilks, A. MacKinnon, A. Offenberger, D. M. Pennington,
	K. Yasuike, A. B. Langdon, B. F. Lasinski, J. Johnson, M. D. Perry, and
	E. M. Campbell, Phys. Rev. Lett. \textbf{85}, 2945 (2000).
	
	\bibitem{Fuchs}
	J. Fuchs, P. Antici, E. D'Humieres, E. Lefebvre, M. Borghesi, E. Brambrink, C. A. Cecchetti, M. Kaluza, V. Malka, M. Manclossi, S. Meyroneinc,
	P. Mora, J. Schreiber, T. Toncian, H. P\'epin, and P. Audebert, Nat.
	Phys. \textbf{2}, 48 (2006).
	
	\bibitem{Ma}
	Y. Y. Ma, Z. M. Sheng, Y. Q. Gu, M. Y. Yu, Y. Yin, F. Q. Shao, T. P. Yu,
	and W. W. Chang, Phys. Plasmas \textbf{16}, 034502 (2009).
	
	\bibitem{Bake}
	M. A. Bake, B. S. Xie, S. Zhang, and H. Y. Wang, Phys. Plasmas \textbf{20}, 033112 (2013).
	
	\bibitem{Yin}
	L. Yin, B. J. Albright, B. M. Hegelich, and J. C. Fernandez, Laser Part.
	Beams \textbf{24}, 291 (2006).
	
	\bibitem{Yin1}
	L. Yin, B. J. Albright, B. M. Hegelich, K. J. Bowers, K. A. Flippo, T. J. T.
	Kwan, and J. C. Fernandez, Phys. Plasmas \textbf{14}, 056706 (2007).
	
	\bibitem{Flippo}
	K. Flippo, B. M. Hegelich, B. J. Albright, L. Yin, D. C. Gautier, S.
	Letzring, M. Schollmeier, J. Schreiber, R. Schulze, and J. C. Fernandez,
	Laser Part. Beams \textbf{25}, 3 (2007).
	
	\bibitem{Silva}
	L. O. Silva, M. Marti, J. R. Davies, R. A. Fonseca, C. Ren, F. Tsung, and
	W. B. Mori, Phys. Rev. Lett. \textbf{92}, 015002 (2004).
	
	\bibitem{Wei}
	M. S. Wei, S. P. D. Mangles, Z. Najmudin, B. Walton, A. Gopal, M. Tatarakis, A. E. Dangor, E. L. Clark, R. G. Evans, S. Fritzler, R. J. Clarke, C. Hernandez-Gomez, D. Neely, W. Mori, M. Tzoufras, and K. Krushelnick,
	Phys. Rev. Lett. \textbf{93}, 155003 (2004).
	
	\bibitem{Esirkepov}
	T. Esirkepov, M. Borghesi, S. V. Bulanov, G. Mourou, and T. Tajima,
	Phys. Rev. Lett. \textbf{92}, 175003 (2004).
	
	\bibitem{Yan}
	X. Q. Yan, C. Lin, Z. M. Sheng, Z. Y. Guo, B. C. Liu, Y. R. Lu, J. X.
	Fang, and J. E. Chen, Phys. Rev. Lett. \textbf{100}, 135003 (2008).
	
	\bibitem{Qiao}
	B. Qiao, M. Zepf, M. Borghesi, and M. Geissler, Phys. Rev. Lett. \textbf{102}, 145002 (2009).
	
	\bibitem{Hong}
	X. R. Hong, B. S. Xie, S. Zhang, H. C. Wu, A. Aimidula, X. Y. Zhao, and
	M. P. Liu, Phys. Plasmas \textbf{17}, 103107 (2010).
	
	\bibitem{Hong2016}
	X. R. Hong, W. J. Zhou, B. S. Xie, Y. Yang, L. Wang, J. M. Tian, R. A. Tang and W. S. Duan,
    \emph{Realization of the radiation pressure acceleration in the near critical density regime}, Phys. Plasmas (under review 2016).
	
	\bibitem{Pegoraro}
	F. Pegoraro, and S. V. Bulanov, Phys. Rev. Lett. \textbf{99}, 065002 (2007).
	
	\bibitem{Yan1}
	X. Q. Yan, M. Chen, Z. M. Sheng, and J. E. Chen, Phys. Plasmas \textbf{16}, 044501 (2009).
	
	\bibitem{Palmer}
	C. A. J. Palmer, J. Schreiber, S. R. Nagel, N. P. Dover, C. Bellei, F. N. Beg, S. Bott, R. J. Clarke, A. E. Dangor, S. M. Hassan, P. Hilz, D. Jung, S. Kneip, S. P. D. Mangles, K. L. Lancaster, A. Rehman, A. P. L. Robinson, C. Spindloe, J. Szerypo, M. Tatarakis, M. Yeung, M. Zepf, and Z. Najmudin
	Phys. Rev. Lett. \textbf{108}, 225002 (2012).
	
	\bibitem{Yin2}
	Y. Yin, W. Yu, M. Y. Yu, A. Lei, X. Q. Yang, H. Xu, and V. K. Senecha,
	Phys. Plasmas \textbf{15}, 093106 (2008).
	
	\bibitem{Chen}
	M. Chen, A. Pukhov, T. P. Yu, and Z. M. Sheng, Phys. Rev. Lett. \textbf{103}, 024801 (2009).
	
	\bibitem{Yu}
	T. P. Yu, A. Pukhov, G. Shvets, and M. Chen, Phys. Rev. Lett. \textbf{105}, 065002 (2010).
	
	\bibitem{Zou}
	D. B. Zou, H. B. Zhuo, T. P. Yu, H. C. Wu, X. H. Yang, F. Q. Shao, Y. Y.
	Ma, Y. Yin, and Z. Y. Ge, Phys. Plasmas \textbf{22}, 023109 (2015).
	
	\bibitem{Kodama}
	R. Kodama, P. A. Norreys, K. Mima, A. E. Dangor, R. G. Evans, H. Fujita,
	Y. Kitagawa, K. Krushelnick, T. Miyakoshi, N. Miyanaga, T. Norimatsu,
	S. J. Rose, T. Shozaki, K. Shigemori, A. Sunahara, M. Tampo, K. A.
	Tanaka, Y. Toyama, T. Yamanaka, and M. Zepf, Nature (London) \textbf{412}, 798 (2001).
	
	\bibitem{Arber}
	T. D. Arber, K. Bennett, C. S. Brady, A. Lawrence-Douglas, M. G. Ramsay, N. J. Sircombe, P. Gillies, R. G. Evans, H. Schmitz, A. R. Bell, and C. P. Ridgers, Plasma Phys. Control. Fusion \textbf{57}, 113001 (2015).
	
	\bibitem{Ma1}
	Y. Y. Ma, Z. M. Sheng, Y. T. Li, W. W. Chang, X. H. Yuan, M. Chen, H.
	C. Wu, J. Zheng, and J. Zhang, Phys. Plasmas \textbf{13}, 110702 (2006).
	
	\bibitem{Kodama1}
	R. Kodama, Y. Sentoku, Z. L. Chen, G. R. Kumar, S. P. Hatchett, Y.
	Toyama, T. E. Cowan, R. R. Freeman, J. Fuchs, Y. Izawa, M. H. Key, Y.
	Kitagawa, K. Kondo, T. Matsuoka, H. Nakamura, M. Nakatsutsumi, P. A.
	Norreys, T. Norimatsu, R. A. Snavely, R. B. Stephens, M. Tampo, K. A. Tanaka, and T. Yabuuchi, Nature (London) \textbf{432}, 1005 (2004).
	
\end{thebibliography}
\end{document}